# Mars' plasma system

## Scientific potential of coordinated multi-point missions:

## "The next generation"

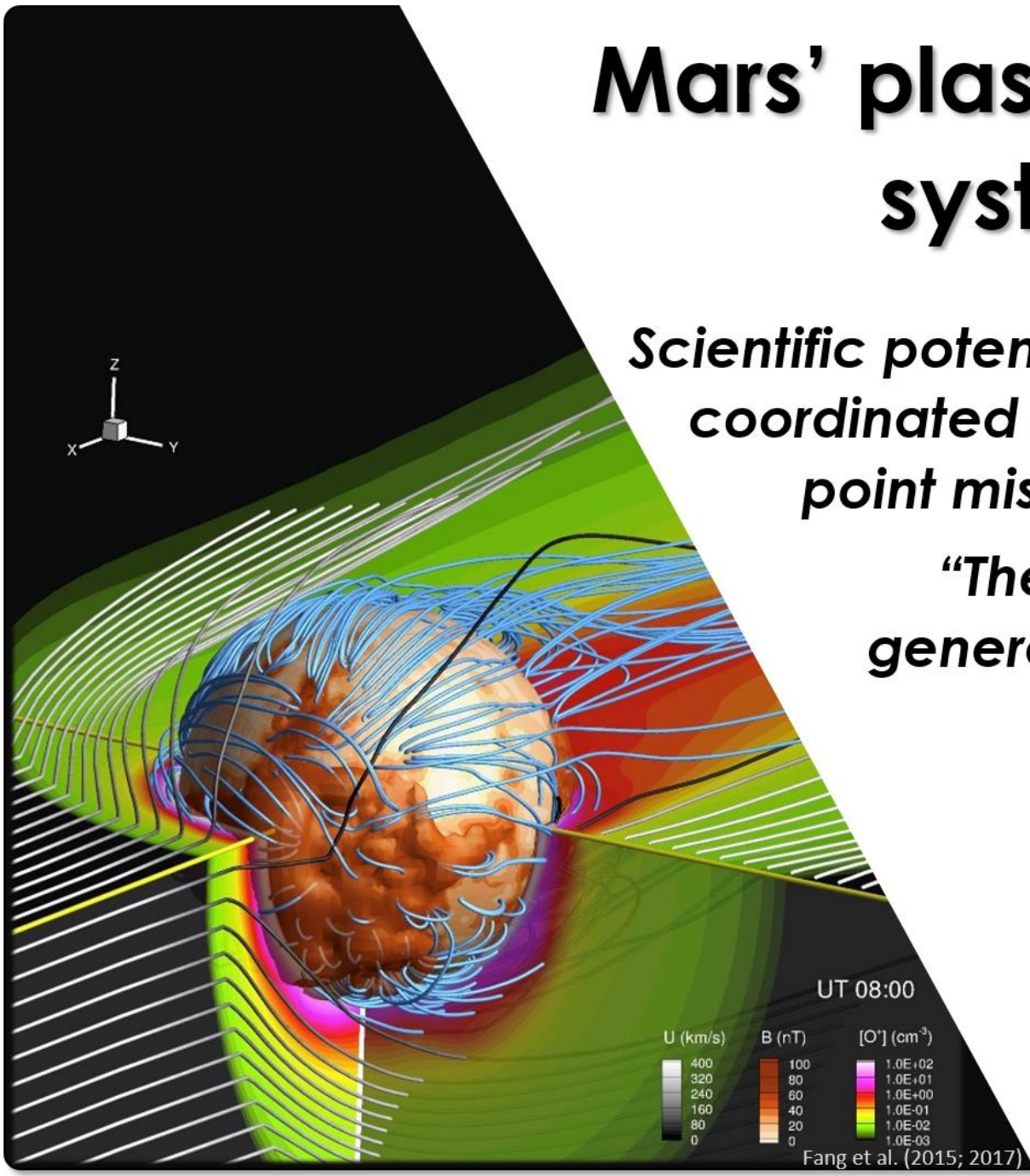

UT 08:00

| U (km/s) | B (nT) | [O⁺] (cm⁻³) |

Fang et al. (2015; 2017)

*A White Paper submitted to ESA's Voyage 2050 Call*

**Contact Scientist:**

**Beatriz Sánchez-Cano**

**Department of Physics and Astronomy,
University of Leicester, United Kingdom**

**bscmdr1 @ leicester.ac.uk        +44 (0) 116 252 3565**

## Executive Summary

The objective of this White Paper submitted to ESA's Voyage 2050 call is to get a more holistic knowledge of the dynamics of the Martian plasma system from its surface up to the undisturbed solar wind outside of the induced magnetosphere. This can only be achieved with coordinated multi-point observations with high temporal resolution as they have the scientific potential to track the whole dynamics of the system (from small to large scales), and they constitute the next generation of Mars' exploration as it happened at Earth few decades ago. This White Paper discusses the key science questions that are still open at Mars and how they could be addressed with coordinated multipoint missions. The main science questions are:

**(i)** How does solar wind driving impact on magnetospheric and ionospheric dynamics?

**(ii)** What is the structure and nature of the tail of Mars' magnetosphere at all scales?

**(iii)** How does the lower atmosphere couple to the upper atmosphere?

**(iv)** Why should we have a permanent in-situ Space Weather monitor at Mars?

Each science question is devoted to a specific plasma region, and includes several specific scientific objectives to study in the coming decades. In addition, two mission concepts are also proposed based on coordinated multi-point science from a constellation of orbiting and ground-based platforms, which focus on understanding and solving the current science gaps.

## Table of Contents

## 1. Motivation and Context

Following the last two decades of near continuous exploration of the Mars' plasma environment, we now know more about the interactions between the different atmospheric layers, and the planetary plasma and the solar wind than any planet other than Earth. Nevertheless, this leaves us with more questions to answer. Thus, the motivation of this White Paper is to demonstrate the key science questions that we are still unanswered at Mars, together with outlaying a mission concept that would answer these questions.

The science questions we propose to answer relate to the fact that the system is strongly coupled in ways which perhaps were unexpected. Each science question is devoted to a specific plasma region. The questions include:

**(i)** How does solar wind driving impact on magnetospheric and ionospheric dynamics?

**(ii)** What is the structure and nature of the tail of Mars' magnetosphere at all scales?

**(iii)** How does the lower atmosphere couple to the upper atmosphere?

**(iv)** Why should we have a permanent in-situ Space Weather monitor at Mars?

In this White Paper, we explore the main scientific aspects that remain unknown at Mars, which are summarized in *Table 1*, and how only simultaneous multi-point observations will help us to solve those scientific questions.

## 1.1. What do we know about the Mars' plasma system?

Unlike most planets in our Solar System, Mars does not have a global magnetic field. The solar wind can interact directly with the upper atmosphere of the planet, and generate an induced magnetosphere (see *Figure 1*). At the subsolar point, this interaction occurs with the ionospheric layer (ion and electron layer at ~100-500 km) (e.g. *Halekas et al., 2017*). However, at larger solar zenith angles (closer to the day-night terminator), the ionosphere is no longer in contact with the solar wind, and a magnetosphere exists in that volume as a layer between heated solar wind plasma flow and the ionosphere (*Vaisberg et al., 2018*). Consequently, the solar wind can strip away Mars' atmosphere very effectively as there is no global magnetic field protecting Mars' atmospheric species (*Jakosky et al., 2015*). In fact, properties of the ionosphere can elucidate the effects of solar wind plasma via structured signatures in the Martian plasma density profiles (e.g., *Withers et al., 2012; Sanchez-Cano et al., 2017; Mayyasi et al., 2018*). The solar wind is, therefore, the outer boundary that controls the Martian plasma system. In addition, Mars has strong magnetic fields at its surface concentrated mostly at a specific region of the southern hemisphere (the so-called crustal fields). These fields can interact directly with the solar wind producing a "hybrid magnetosphere" in that region, i.e. with features of both induced and intrinsic magnetospheres, that changes as the crustal magnetic fields rotate with the planet (e.g. *Ma et al., 2014*) (see *Figures 1 and 2*). This magnetic environment, coupled with electric fields from multiple sources (e.g. *Dubinin et al., 2008; Lillis et al., 2018*) determines the ion and electron motions and hence whether they escape, precipitate at low energies to be reabsorbed, or at high energies (>~1 keV) to cause sputtering escape of neutrals (*Wang et al., 2014*). Moreover, crustal magnetic fields play an important role in guiding plasma motion, such as a large hemispheric asymmetry in the magnetosphere, ionosphere, and the density of escaping ions (*e.g. Vaisberg et al., 2018*). On the other hand, Mars has strong lower atmospheric cycles such as the water or $CO_2$ cycles (e.g. *Smith et al., 1999*), as well as global dust storms (e.g. *Montabone et al., 2015*) and gravity waves (e.g. *Yigit et al., 2015; England et al., 2017; Terrada et al., 2017*), that are produced by different phenomena related mainly to the low gravity of the planet, its extreme topographic features, and its large orbital ellipticity. These lower atmospheric phenomena can at times drive the behaviour of the ionosphere. In summary, **the Martian space environment is a Complex System with simultaneous downward and upward couplings, which need to be understood.**

## 1.2. Scientific potential of coordinated multi-point observations

Our experience from 60 years of space exploration at Earth tells us that we need simultaneous multi-point observations of the whole Martian system in order to gain an adequate understanding of Mars as a dynamic system. Such multi-point missions have revolutionized the understanding of the terrestrial solar wind-magnetosphere-ionosphere coupling, like for example, with the Cluster-II (*Escoubet et al., 2000*), THEMIS (Time History of Events and Macroscale Interactions during Substorms) (*Angelopoulos, 2008*), Swarm (*Olsen et al., 2013*), and MMS (Magnetospheric Multiscale Mission) (*Burch et al., 2015*) missions. At Mars, prototype multi-spacecraft studies have been completed, e.g. Mars Express-Mars Global Surveyor, Mars Express-Rosetta, and now Mars Express-MAVEN (Mars Atmosphere and Volatile EvolutioN), together with studies using data from Mars Reconnaissance Orbiter (e.g. radar), Mars Odyssey (e.g. neutron monitor), and Mars Science Laboratory (e.g. radiation monitor). Nevertheless, **better-coordinated multi-spacecraft studies with high temporal resolution will enable the questions posed here to be answered.** In other words, only **multiple and simultaneous observations** at different parts of the Martian plasma system will unravel the key mechanisms that make Mars a unique system, strongly coupling its surface, lower, mid and upper atmosphere, ionosphere, exosphere, induced magnetosphere and the solar wind (*Figure 1*). **This will allow us to understand, e.g. spatial versus temporal effects, small scale disturbances, flow of energy and mass through the system, and the response of the downstream system to changes in the upstream solar wind.**

The last two decades have seen a significant increase in the amount and variety of observations characterizing the thermal structure and basic composition of Mars' atmosphere, from the surface to the exosphere. It also has opened the door to the understanding of the physical processes that control the current Martian climate, from the general circulation, to the role of photochemistry, clouds, development of dust storms (both local and global), and channels and rates of atmospheric escape. However, despite this progress, we do not understand many of the physical processes that drive matter and energy flow between and within the various atmospheric reservoirs yet. For example, we do not know how low atmospheric cycle effects propagate towards the upper atmosphere and contribute to enhancing escape processes yet. In addition, **there are still two important observational gaps in the Martian system that no mission has been able to fully explore**: **the 3D structure of the full Martian tail and its dynamics, and the lower Martian ionosphere from the surface to ~80 km (**which has only been sampled during the descent of the 2 Viking landers (*Hanson et al., 1977; Hanson and Mantas, 1988*).

***Table 1:*** *Summary of the Main Science Questions and Specific Scientific Objectives*

| in Science Questions | Specific Scientific Objectives *(per section)* |
|---|---|
| **SCIENCE QUESTION 1** <br><br> **How does solar wind driving impact on magnetospheric and ionospheric dynamics?** <br> *(section 2.1)* | 2.1.1. How are the Martian induced magnetosphere and its plasma boundaries affected by solar wind variability? <br> 2.1.2. How is the Mars-solar wind interaction affected by the coupling with the crustal magnetic fields? <br> 2.1.3. How are the current systems at Mars driven by the solar wind - planet interaction?? <br> 2.1.4. The mystery of the energy budget at Mars: solar wind ionospheric heating <br> 2.1.5. Can the solar wind enhance the neutral and ion escape rates? |
| **SCIENCE QUESTION 2** <br><br> **What is the structure and nature of the tail of Mars' magnetosphere at all scales?** <br> *(section 2.2)* | 2.2.1. What is the large scale structure of the Martian tail, and does magnetic reconnection occur there? What are the plasma sheet dynamics and how do they vary with solar activity? <br> 2.2.2. How efficient is plasma transported and to where in the nightside and at different solar activity levels? <br> 2.2.3. What is the physical mechanism that explains nightside precipitation (and auroras) in regions far from magnetic fields? |
| **SCIENCE QUESTION 3** <br><br> **How does the lower atmosphere couple to the upper atmosphere?** <br> *(section 2.3)* | 2.3.1. What is the structure of the day and nighttime ionosphere (including the bottomside ionosphere)? <br> 2.3.2. Does plasma reach the Martian surface? <br> 2.3.3. Quantitatively, what is the role of lower atmospheric effects on the ionosphere? <br> 2.3.4. To what extent does the ionosphere permit and inhibit radio communication at the surface? <br> 2.3.5. What role do winds play on wave propagation? <br> 2.3.6. What are the roles of small scale ionospheric irregularities and electrodynamics in the Martian ionosphere? <br> 2.3.7. How do low atmospheric cycles affect the upper atmosphere and escape? |
| **SCIENCE QUESTION 4** - **Why should we have a permanent in-situ Space Weather monitor at Mars?** *(section 2.4)* ||

Another important aspect recently discovered is the significance of the vertical coupling between the Mars' atmosphere and plasma systems. New evidence demonstrates that different regions of the Martian atmosphere are fundamentally interconnected, and behave as a unique and coherent system (e.g. *Bougher et al., 2015, 2017; Jakosky, 2015; Montmessin et al., 2017; Sanchez-Cano et al., 2018a*). This means that the whole atmospheric structure reacts together to external and internal sources of variability, and therefore, plays an important role in the volatile escape processes that have dehydrated Mars over the Solar System's history, holding clues to the evolution of Mars' climate. Comparative studies at Earth and Mars have demonstrated that such coupling can be driven from above the system and below (e.g. *Figure 1* for the Mars case) (e.g. *Mendillo et al., 2003; 2018*). At Mars, this is a growing topic, although **still at a preliminary stage thanks to missions such as Mars Express and MAVEN, but requiring a longer and more exhaustive global coverage of observations.**

The importance of continuous Space Weather observations is also currently being uncovered, especially in preparation for the future human exploration of Mars where communications between surface and orbiters is essential. For example, solar storms are sources of very intense short-term planetary variability, whose effects are possibly hazardous as they enhance auroras, create large radiation showers via energetic particle precipitation into the atmosphere, can produce technologies disruptions, and play a very important role in atmospheric escape processes, which are currently a major research topic in Mars' exploration (e.g. *Jakosky et al., 2018*). All these effects speak of the real need of having continuous Space Weather observations at different Solar System positions, and in particular at Mars, where **an efficient and continuous thermosphere – ionosphere – magnetosphere –solar wind monitoring service is needed in the eve of the Martian human exploration.**

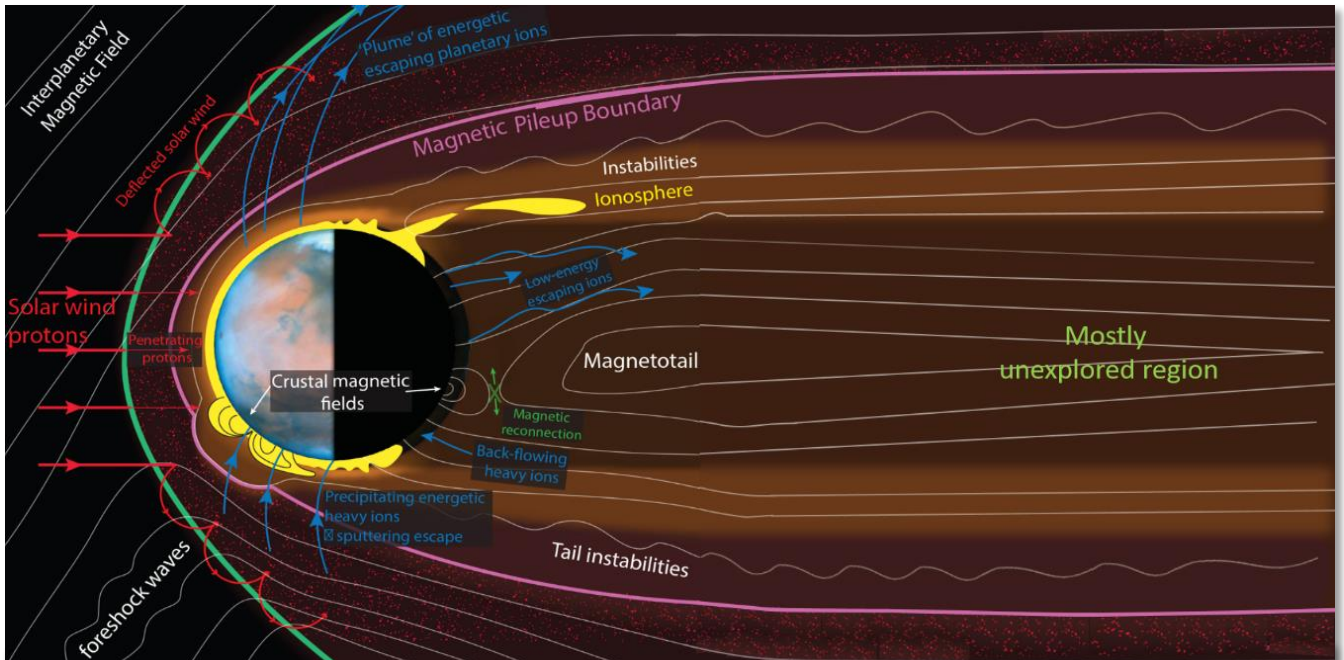

*Figure 1: Mars' plasma system scheme showing the main physical processes known to occur at Mars. The Sun is to the left. Multi-point plasma measurements are needed to understand the whole dynamic system at Mars. (Picture adapted from Lillis et al. (2019), adapted from Fran Bagenal and Steve Bartlett (CU-LASP)).*

Understanding how each planet, moon and comet respond to Space Weather variability is an important task that gives us great information on the evolution of the solar wind and solar wind transient structures (i.e. solar storms), and how different magnetized/un-magnetized environments (based on the presence/lack of an intrinsic dipole field) react to different energy inputs from the solar wind. **We now have a unique opportunity at Mars** to perform **comparative planetology science,** which will allow us to extrapolate knowledge from one planet to another (including exoplanets), and forecast adequate planetary responses with more accuracy, like for example to Space Weather events. Moreover, it will help to assess the possible habitability of planets and their moons. Understanding the effects of the variable solar wind as well as of the intrinsic Mars' variability on the Mars' plasma system requires simultaneous measurements of the properties of both Martian and solar wind plasma.

## 2. Science Questions and Objectives

In this section, we develop one by one the main science questions and specific scientific objectives that we consider should be the object of study in the coming decades. The objective is to provide a more holistic knowledge of the dynamics of the Martian plasma system from its surface up to the undisturbed solar wind outside of the induced magnetosphere. It is divided into four main blocks that account for different regions of the plasma system, and are summarized in *Table 1*.

### 2.1. SCIENCE QUESTION 1

> **How does solar wind driving impact on magnetospheric and ionospheric dynamics?**

Despite the importance of the continuous plasma observations since the 90s from Mars Global Surveyor (MGS), Mars Express and recently MAVEN, we still lack a clear characterisation of how solar wind dynamics drive the magnetosphere and ionosphere. This includes the behaviour and formation of all plasma boundaries, the actual role of crustal magnetic fields on the whole system, solar wind heating effects which recently have been revealed to be much more important than anticipated, and escape processes. In this section, we develop each of these topics in detail.

### 2.1.1. How are the Martian induced magnetosphere and its plasma boundaries affected by solar wind variability?

For a traditional magnetosphere, the magnetopause is its outer boundary. This is the boundary that separates the region dominated by the planetary magnetic field from the region dominated by the solar wind. However, this definition is different for unmagnetized bodies like Venus, Mars or comets because they do not have a global intrinsic magnetic field, and their interaction with the solar wind occurs at their upper atmospheres. In these cases, the solar wind induces a magnetosphere, which is found at a much closer distance than at magnetized planets such as Earth (e.g. *Bertucci et al., 2011*). These outer boundaries are usually referred to as Magnetic Pile-up Boundary (MPB), Ion Composition Boundary (ICB) or Induced Magnetosphere Boundary (IMB) (e.g., *Nagy et al. 2004, Matsunaga et al., 2017, Halekas et al., 2018; Espley et al., 2018*). Moreover, it is not clear whether the ionopause (a tangential discontinuity in the ionospheric thermal plasma density that marks the end on the ionosphere) (*Schunk and Nagy, 2009*) is somehow related to any of those boundaries. The main reason for the lack of a common definition is the limited plasma instrumentation available on the earlier Mars' missions, which resulted in most boundaries being defined based on only one or two measurement types. It was not until the MAVEN mission arrived at Mars in 2014 carrying a comprehensive plasma and magnetic field instrumentation, that the various boundaries could start to be studied in detail (see e.g., *Matsunaga et al. 2017, Holmberg et al. 2019*).

MAVEN is shedding light on many of our questions regarding the Martian system, but detailed magnetospheric observations raised many new questions. One current discussion concerns the relevance of the various boundaries acting as the outer boundary of the induced magnetosphere, the relationship between the boundaries and their dependence on factors such as the solar wind dynamic pressure, interplanetary magnetic field (IMF) strength and direction, solar extreme-ultraviolet (EUV) flux, and crustal magnetic field strength (e.g., *Edberg et al., 2009, Xu et al., 2016, Matsunaga et al., 2017, Halekas et al., 2018, Holmberg et al., 2019*). This means that **the structure of the Martian magnetosphere is still not fully characterized** and the parameters determining the structure **are not conclusively verified or quantified**.

All current missions lack a crucial component for studying magnetospheric structures and dynamics: they **do not have a continuous solar wind monitor**. For example, when studying how a magnetospheric structure varies with changes in the solar wind, a single spacecraft measurement lacks simultaneous solar wind measurement and has to rely on solar wind models that are subject to at times significant uncertainties, especially during Space Weather events (*Ruhunusirl et al., 2018; Hurley et al., 2018; Romanelli et al. (2018); Dong et al., 2019*). A single spacecraft measurement cannot disentangle spatial versus temporal variations of magnetospheric structure. Hence, it is easy to conclude that when studying global structures that exhibit both temporal and spatial variations, **single point measurements have a high risk of providing erroneous results**. A multi-spacecraft mission provides the possibility to simultaneously measure changes in the solar wind and to record the magnetospheric response at multiple locations in the Martian induced magnetosphere. Such a mission **would be crucial in finally revealing the true nature and flow of energy within the Martian induced magnetosphere**. It would also be important in understanding the structures of induced magnetospheres in general. Even though the concept of an induced magnetosphere might seem simple at first, studies of the different induced magnetospheres in our Solar System, have shown **a more complex interaction** than previously expected. A holistic analysis of the Martian induced magnetosphere would also be very useful for comparative studies of the induced magnetospheres of Venus, comets and moons and how they couple with their ionospheres, teaching us more about how our Solar System works.

### 2.1.2. How is the Mars-solar wind interaction affected by the coupling with the crustal magnetic fields?

Mars is unique among the terrestrial planets in that it has no strong intrinsic dipole magnetic field to protect its atmosphere/ionosphere from the impinging solar wind, but does have highly non-uniformly distributed and locally strong crustal magnetic sources (*e.g. Connerney et al., 2001*). Despite being studied for two decades since their discovery, **the role of the crustal fields in driving and disturbing the near-Mars space environment**, on both global and local scales, is **still not well understood**. It is known (more on a statistical sense) that crustal magnetism exerts an important control on surrounding ionospheric properties, locations of plasma boundaries (including ionopause, magnetic pileup boundary, photoelectron boundary, and even the bow shock) (e.g. *Mitchell et al., 2001; Hall et al., 2016; Garnier et al., 2017*), and magnetospheric configurations (including magnetic field topologies and structure of the magnetotail current sheet) (e.g. *Weber et al., 2019*). In addition, it has been recently revealed that the crustal field not only has a shielding effect against atmospheric loss due to the solar wind stripping, but also an opposite escape-fostering effect by regulating the day-to-night transport and the overall net effect of these crustal anomalies on ionospheric escape is still not known (*Fang et al., 2017*). There are also important small-scale effects in association with the crustal field over cusp regions, such as particle and wave penetration, field-aligned currents, and ionospheric electrodynamics and large scale ionospheric perturbations (e.g. *Matta et al., 2015; Andrews et al., 2018*). Another complexity comes from temporal variations due to the continuous rotation of the planet and thus the ever-changing crustal field orientation to the Sun (e.g. *Fang et al., 2015, 2017*). This can be seen in the simulation presented in *Figure 2*. This Figure shows a global time-dependent MHD simulation of the interaction of the steady solar wind (white-grey colours) with the Martian plasma system, where the intensity of the crustal field on the Martian surface is represented in shades of brown, and the Martian O⁺ density in green-purple scale colours. The simulation allows the crustal magnetic fields to rotate with time (field lines in shades of blue). As can be seen, the magnetic topology with respect to the solar wind changes dramatically in only half day due to the rotation of the planet. This situation is even more complex when Space Weather events hit Mars.

The availability of ever increasing spatial and temporal coverage of space-borne satellite observations and recent numerical modelling advances have significantly broadened and deepened our understanding of the interaction between Mars and the solar wind. However, **detailed and quantitative descriptions are still missing on the role of the crustal field in the mass and energy flow throughout the ionosphere and magnetosphere** as well as particle and energy exchange particularly over cusp regions. These challenges **require multi-point observations** covering upstream drivers and downstream responses and relating activities and variabilities among different space elements in the Mars' system. It is important to improve global-scale and local-scale model development, in which the distributions of neutral and charged particles and the electromagnetic field are self-consistently accounted for. The advances in modelling and data integration are a key factor to solve this long-standing science question, which in turn have a very significant effect on the rest of the system. There is a clear need for accurate global models, based on multi point measurements, which are crucial for model validation. **Modelling and data analysis approaches need to be extensively tested and integrated into a coherent picture of the Mars-solar wind interaction from a system perspective**.

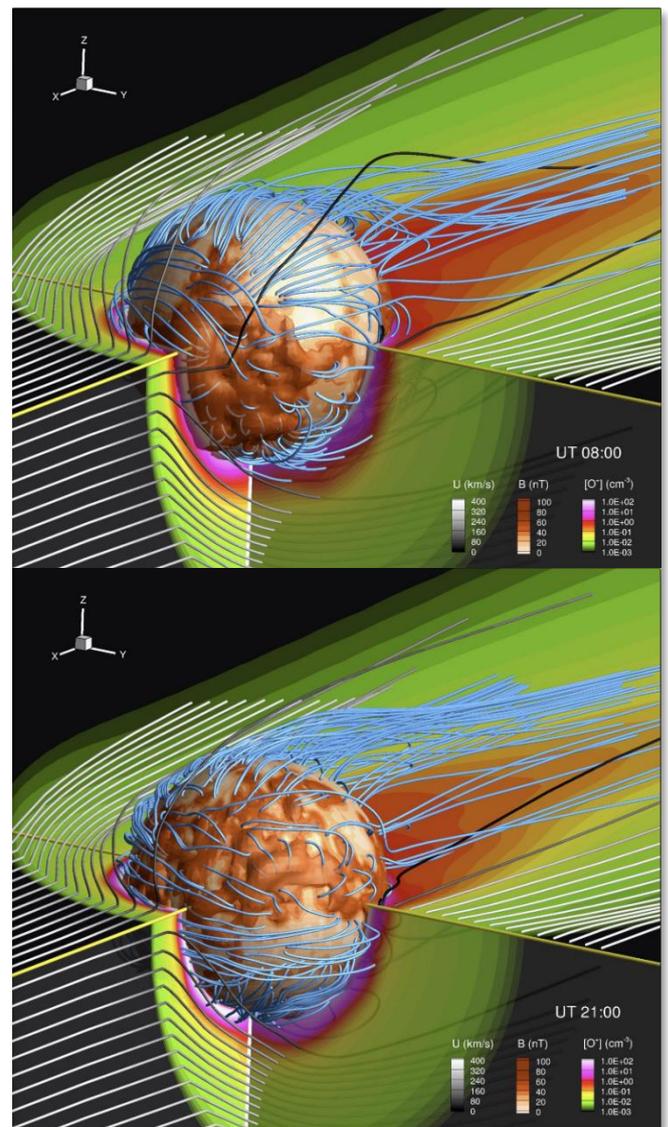

*Figure 2: Global time-dependent MHD simulation of the Mars-solar wind interaction under quiescent solar wind conditions but allowing the crustal magnetic fields to continuously rotate with time. Top panels: at 8:00 UT. Bottom panel: at 21:00 UT (more details in Fang et al. 2015, 2017).*

### 2.1.3. How are the current systems at Mars driven by the solar wind - planet interaction?

Planetary current systems are a natural connection between different regimes within a planet system. This means that different regions of a planet with different plasma populations, such as the solar wind, the magnetosphere, the ionosphere, and the ground, are frequently linked by currents.

The bow shock is the first place where the supersonic solar wind starts interacting with a planetary obstacle, as it decelerates the incoming solar wind and compresses the magnetic field in the magnetosheath region, so that the plasma can flow around the obstacle behind. At Earth, it has been realized only very recently that the currents in the bow shock logically connect to other regions of diverging currents in the magnetospheric-ionospheric systems, and that under certain circumstances, it is the main generator of the entire solar wind – bow shock – magnetosphere system at Earth (*López et al., 2011*). At Mars, however, the global current system is unknown, although assumed to be somehow qualitatively similar to Earth. Mars has a complex magnetic topology however (see *Figure 2*), and the ionospheric current signatures are far from well understood as there are only very few measurements (*Fillingim, 2018*). Moreover, although we have a generally good knowledge of the basics of the Martian bow shock and MPB (i.e., average location, how it responds in general to changes in the solar wind, etc) (*e.g. Mazelle et al., 2004, Gruesbeck et al., 2018, Hall et al., 2016; 2019*), **we do not know the detailed and local physics of the bow shock** (e.g. *Meziane et al., 2017; 2019; Mazelle et al., 2018*), **as well as it is not yet understood in the context of current systems**.

Therefore, there is a clear need for investigations of the variability of the bow shock and subsequent currents with solar wind and solar activity variations, which cannot be carried out with current instrumentation. **Higher cadence measurements and multi-point measurements,** such as MMS at Earth, **are required to qualitatively evaluate these current systems**. Necessary studies include understanding the variation of these currents with heliocentric changes heliocentric changes

(which affect the amount of solar radiation and solar wind that reach Mars) and different solar cycle phases, as well as different current divergences and their connection to the Martian induced magnetosphere, other than being an optional by-product in most MHD models.

### 2.1.4. The mystery of the energy budget at Mars: solar wind ionospheric heating

The typical plasma length scales within the Mars induced magnetosphere are similar to the solar wind standoff distance and it is expected to lead to the direct transfer of energy between the solar wind and ionosphere (e.g. *Moses et al., 1988*). Such processes may play an important role in the energization of the ionosphere and subsequent escape to space, particularly in the past, when the Sun is thought to have been more active, leading to a stronger solar wind – Mars interaction.

**The energy budget at Mars is not sustained from solar heating alone** (*Matta et al., 2014). Figure 3* shows an example where only when an additional topside ion heating flux is included in a numerical simulation (in this case for $O_2^+$), the resulting topside $O_2^+$ profile temperatures increase being able to reproduce observations (*Matta et al., 2014*). The Mars Express and MGS missions have observed this solar wind – planet interaction, but **limitations on spacecraft orbits and/or instrumentation have meant that only glimpses of this energy transfer have been observed** (*Lundin et al., 2004, Barabash et al,., 2007*). More recent observations by the MAVEN spacecraft have built upon these earlier studies. Compressive, magnetosonic waves generated in the foreshock region have been observed to propagate into the dayside ionosphere and heat the ionosphere via stochastic heating due to the non-conservation of the magnetic adiabatic invariant (*Collinson et al., 2017; Fowler et al., 2018*). Moreover, ongoing studies are also showing that plasma temperatures in the upper atmosphere of Mars can only be reproduced when additional external heating is provided to the system. The solar wind is an ideal candidate for such energy deposition as it can produce as well as heat ambient plasma to values that are consistent with measurements. Wave heating can become important at high altitudes near the top of the ionosphere (e.g. *Ergun et al., 2006*), while ionospheric ions are heated most predominantly via collisions with electrons at low altitude. Yet, **those interactions alone cannot explain the observed ion temperatures.** The solar wind could also explain such a discrepancy. **Ion temperature measurements are** presently **mostly lacking** at Mars, with only two measurements made with the Viking Lander RPAs (*Hanson et al., 1977*), and very few retrievals from MAVEN (*Fowler et al., 2018*).

**Simultaneous ionospheric and electron temperatures at multiple stages** in the system (i.e. bow shock, magnetosheath, upper and lower ionosphere) **are absolutely needed** to fully understand and explain the energy budget conundrum at Mars, including to **start to understand how energy flows from the top to bottom of the system**. The nature of single point measurements make it difficult to quantify the time versus spatial evolution of such heating events, and only provide a limited snapshot of the heating region. **Multi-point measurements will be crucial to unravelling how energy flows from the solar wind into the ionosphere.** Magnetic field and plasma moments will be required at cadences able to resolve fundamental plasma time scales (such as the ion cyclotron frequencies) to quantify this energy transfer. Measurements will need to span both the thermal and superthermal energy ranges.

### 2.1.5. Can the solar wind enhance the neutral and ion escape rates?

Martian atmospheric losses are mainly led by thermal escape of neutral hydrogen and photochemical escape of neutral oxygen. These mechanisms, together with ion outflow, sputtering, and pickup ion escape, are believed to have led to the disappearance of liquid water on Mars (e.g. *Jakoski, 2015; Chassefiere and Leblanc, 2014*). However, **direct measurements of the escaping neutral hydrogen and oxygen atoms is impossible with current technology** due to the low density and energy of escaping neutrals and only theoretical and indirect estimations can be done.

Regarding water-species, the solar wind effects on atmospheric loss is beginning to be examined with MGS, Mars Express and MAVEN, and Space Weather events have been shown to greatly enhance the escape rate of water-originating species from Mars (e.g., *Lundin et al., 2008; Futaana et al., 2008; Edberg et al., 2010; Opgenoorth et al., 2013; Ma et al., 2014; 2017; Jakosky et al., 2015a; Curry et al., 2015; Luhmann et al., 2017; Mayyasi et al., 2018; Fowler et al., 2019*). However, **in situ ionospheric observations are limited** to a single swath every few hours from these missions. Although the latitude and local time coverages of these various missions differ, **individual spacecraft measurements still make it difficult to determine the large scale response of the ionosphere to dynamic space weather events.**

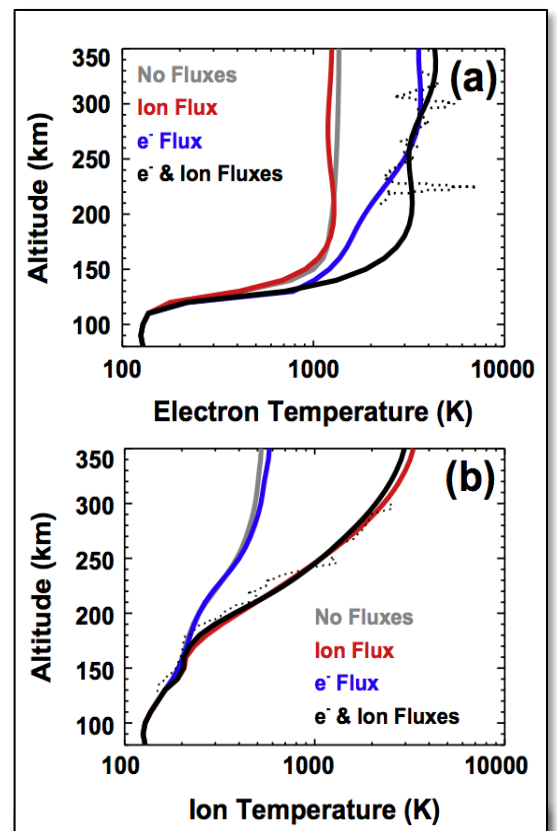

## 2.2. SCIENCE QUESTION 2

### What is the structure and nature of the tail of Mars' magnetosphere at all scales?

The Martian magnetosphere and ionosphere nightside are only now starting to be untangled thanks to the MGS, Mars Express and MAVEN missions. We know significant structure and variability exists in both the dayside and the nightside parts of the system. However, **we do not know** the full implications of this variability because one of the **main aspects of a planetary system that still remains unknown** at Mars is the length and main characteristics of the Martian tail, as well as its dynamics (*Figure 1*). Although most of the missions have visited the Martian nightside, **none of them has travelled deep enough** (>3-4 Mars radii), with the only exception of few transits from Mars 4, and Mars 5 (*Vaisberg et al., 1976; Vaisberg and Smirnov, 1986*), in order to perceive where

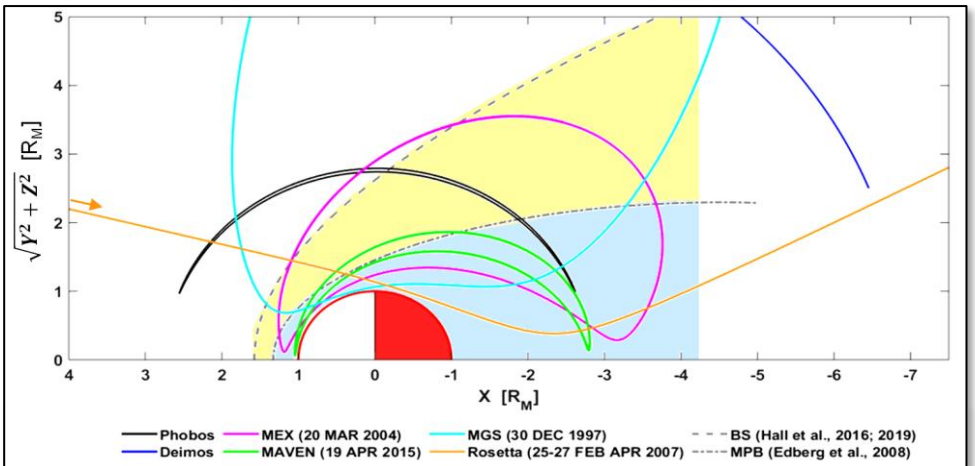

**Figure 4:** *Orbit configuration in MSO-cylindrical coordinates of the different missions that have transited the Martian tail with specific plasma instrumentation. The orbits correspond to their further transit within the Martian tail. As observed, the tail from ~3.5-4 Mars radii has not been much explored, with the exception of the Rosetta and Mars-4,5 single flybys to Mars. We note Phobos-2 and Mars-4,5 are not included in this figure but discussed. The Sun is to the left. BS an MPB stand for bow shock and magnetic pilled-up boundary respectively. Phobos and Deimos orbits are plotted for context.*

Recent MAVEN data together with modelling observations suggested that magnetic reconnection occurs in the Martian tail on a similar fashion to what happens at Earth. However, in Mars' case, an additional cause may also be reconnection of the IMF with the crustal fields (*DiBraccio et al., 2017*). In addition, similar signatures to substorms at Earth have been observed, as well as plasma sheet flapping and high-energy planetary ions ($O^+$ and $O_2^+$) escaping within the current sheet (*DiBraccio et al., 2017*). These crustal fields also have some significant effects over the global escape rate, as recently demonstrated with long-term observations from Mars Express (*Ramstad et al., 2016*).

the tail terminates and what dynamics are present (*Figure 4*). Understanding how the **whole system** (including the far tail) behaves is essential for ion outflow and inflow processes (particle precipitation), as well as to assess the 3D structure and life-time of the different dynamic processes, for which **our current knowledge is very limited**.

### 2.2.1. What is the large scale structure of the Martian tail, and does magnetic reconnection occur there? What are the plasma sheet dynamics and how do they vary with solar activity?

Our knowledge of the magnetospheric tail is mainly based on magnetic observations from MGS and MAVEN, and particle observations from Mars Express and MAVEN. In addition, the Rosetta mission did a single flyby to Mars in its way to comet 67P/ Churyumov-Gerasimenko that allowed us to get more knowledge of the Martian plasma system (*Edberg et al., 2009*) (*see Figure 4*). In general, it seems that Mars dayside ionosphere exerts significant control over the nightside-induced magnetosphere. Early observations in the 70s estimated that the Martian tail diameter (normalized by the planet's radius) appeared to be about twice as large as the width of the Venus' induced magnetotail, which was an indication of evidence for the presence of an intrinsic global magnetic field (e.g. *Vaisberg and Smirnov, 1986*), which has since been shown to not be the case.

The solar activity also seems to play a role in the structure and variability of the tail, like during solar maximum conditions when a Venus-like tail configuration with the current sheet shifted to the dawnside direction is found. On the contrary, solar minimum conditions result in a flipped tail configuration with the current sheet shifted to the duskside direction (*Liemohn et al., 2017*). Moreover, the lack of observations at further distances create enormous uncertainties on the location of the different plasma boundaries, which gradually becomes significant down the tail (*Fang et al., 2017*). Understanding these variations has an important implication for the amount of integrated tailward escaping ions (e.g. *Fang et al., 2015; Garnier et al., 2018*).

Evidence clearly indicates that the Martian tail is very active and different from other planetary magnetic tails and comet tails. However, **we need missions that systematically transit the Martian tail far from the planet together with simultaneous solar wind observations** in order to understand and observe the behaviour of the tail, its length, and understand whether tail reconnection similar to Earth's tail (and substorms) systematically occurs. Bulk plasma escape in the form of tailward traveling plasmoids have been observed at Mars (e.g. *Brain et al., 2010*), however, observational limitations mean that a full characterization of these plasmoids has thus far been unobtainable, something that would be remedied with a dedicated magnetotail mission. Moreover, differences in solar activity/solar wind should play notable roles in the dynamics of the tail, specifically on the plasma sheet. **Therefore, monitoring of the tail and of the solar wind for a whole solar cycle is needed.**

### 2.2.2. How efficient is plasma transported and to where in the nightside and at different solar activity levels?

The nightside Martian ionosphere near the terminator is more complicated than in principle expected, especially below 300 km. In addition to partial photoionization (at high altitudes where light is still present beyond the terminator) (e.g. *Němec et al., 2015*) and electron impact ionization (e.g. *Girazian et al., 2017*), day-to-night plasma transport is also an important source of ionization (e.g., *Duru et al., 2011; Němec et al., 2010; Withers et al., 2012; Girazian et al., 2017*), being dominant over solar wind electron precipitation for about 5,000 s after terminator crossing (*Cui et al., 2015*). A similar process is known to occur on other terrestrial planets such as Venus (*Knudsen et al.,1980; Spenner et al., 1981*) and Titan (*Cui et al., 2009, 2010*). However at Mars, transport has been discovered to not be symmetric between hemispheres, having notable dawn-dusk and north-south asymmetries and varying among different ion species (*Cao et al., 2019*). As for many other processes, crustal magnetic fields seem to be the responsible source for such anomalous behaviour.

Despite various studies focusing on the variability and the driving force of the nightside Martian ionosphere near the terminator, **it is unclear how such a transition region is affected by the ambient crustal magnetic fields**. These fields are known to cause large variability in both day and nightsides (*Němec et al., 2015*). They seem to shield precipitating electrons and suppress the day-to-night transport (*Cao et al., 2019*). However, **their full dynamic role on plasma transport at the terminator is not fully understood yet**. Moreover, another important factor to consider is that **long-term observations of plasma transport at the terminator are needed in order to understand if the solar cycle plays a role** there, and if so, quantify it at the different Martian hemispheres (dawn/dusk, south/north). This is important for also understanding the long-term variability of several escape processes. Thanks to the 15 years or so of Mars Express ionospheric observations, we know now that the solar cycle together with Mars' heliocentric distance are major driving mechanisms in Mars' ionosphere variability (*Sanchez-Cano et al., 2015b, 2016a*). Therefore, **it is expected that plasma transport to the nightside has also**

**a strong dependence with solar cycle**, although their importance at the different Martian hemispheres **needs still to be quantified.**

### 2.2.3. What is the physical mechanism that explains nightside precipitation (and auroras) in regions far from magnetic fields?

On the deep-nightside (close to midnight), electron precipitation is usually the dominant source of energy input to the Martian atmosphere (*Lillis and Brain, 2013*), especially over regions of closed crustal magnetic fields lines (e.g. *Lillis et al. 2018; Němec et al., 2015*). Thanks mainly to the MAVEN mission, we now know that electron precipitation occur everywhere on the Martian nightside. For example, it has revealed that diffuse aurora can be seen at any location on the Martian nightside when a solar storm impacts Mars. These auroras emissions are known to be caused by solar energetic particles (SEPs), specifically electrons accelerated to energies of ~100 keV at the Sun and heliospheric shock fronts (*Schneider et al., 2015; 2018*). Also, the same Space Weather phenomenon is known to create low ionospheric layers (below 100 km) everywhere over the nightside after SEP electrons ionize the very low atmosphere, producing multiple radar and operation difficulties for several days (*Sánchez-Cano et al., 2019*). Therefore, electron precipitation on the deep-nightside is not an isolated effect.

We still need to understand why these energetic particles from the solar wind end up impacting on the nightside atmosphere of Mars, far from the regions where crustal magnetic fields are. In other words, *how do those electrons reach that part of the atmosphere?* At Earth, this phenomenon is explained by magnetospheric tail reconnection during which charged particles travel along closed magnetic field lines into the Earth's atmosphere (*Dungey, 1961*). However at Mars, **such a mechanism has not been confirmed, and perhaps may be related to the still little-known processes that occur within the far tail** (see *Figure 1, and Section 2.2.1*).

## 2.3. SCIENCE QUESTION 3

> ### How does the lower atmosphere couple to the upper atmosphere?

Measurements made of the structure of Mars' ionosphere from orbital platforms are well in advance of all other planetary bodies in the Solar System with the exception of Earth. Current understanding of Mars' ionosphere and thermosphere is largely informed by "top-down" observations, i.e. those made from spacecraft in orbit, in contrast to the manner in which our understanding of Earth's ionosphere developed. Indeed, **no measurements of the Martian ionosphere have been made from the surface at low radio frequencies.** Consequently, our knowledge of the lower ionosphere of Mars is largely informed by measurements from orbit, combined with theoretical modelling, and **significant gaps are present in our knowledge.** Understanding of both the structure and dynamics of the lower ionosphere, and its coupling with the neutral atmosphere, could be greatly advanced using ground-based measurements.

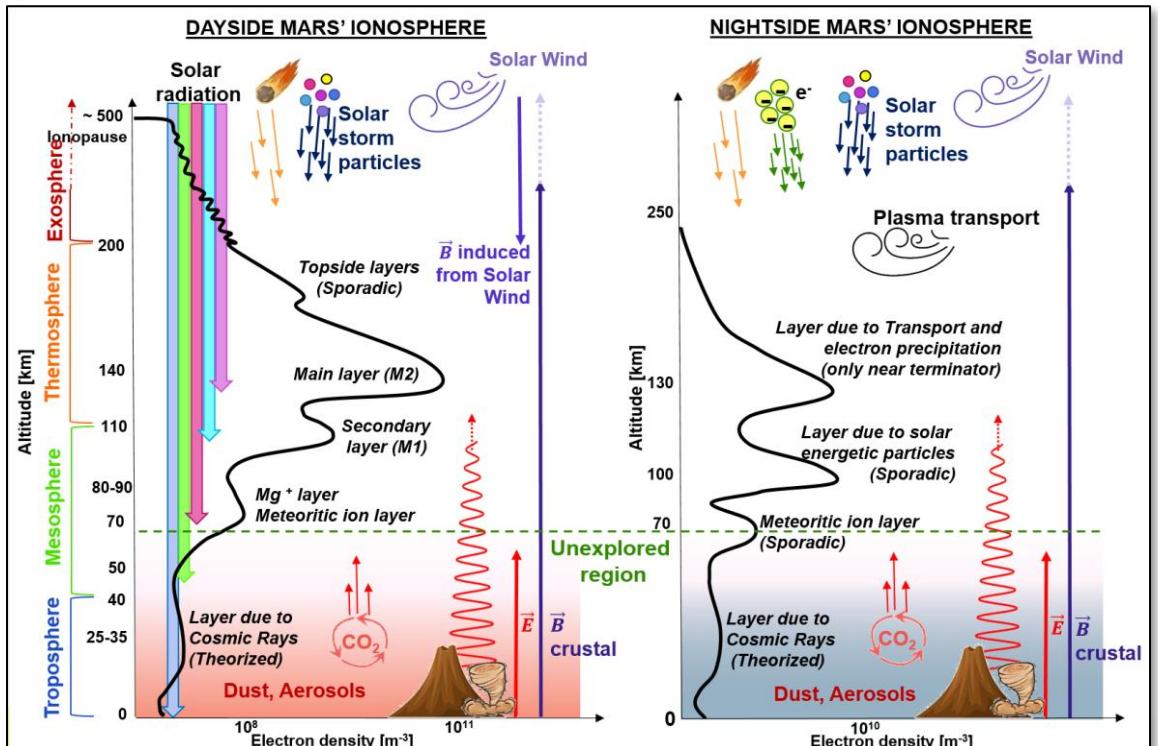

**Figure 5:** *Typical dayside (left) and nightside (right) Martian ionospheric profiles. The different atmospheric layers, and main internal and external forcings: solar radiation, meteors, electron and solar storm particle precipitation, solar wind, magnetic fields, gravity waves, dust storms, and atmospheric cycles are also indicated. Figure from Beatriz Sánchez-Cano (University of Leicester).*

### 2.3.1. What is the structure of the day and nighttime ionosphere (including the bottomside ionosphere)?

The dayside ionosphere of Mars is mainly formed by photoionization of the $CO_2$ dominated atmosphere by a combination of solar EUV, X-ray radiation and photoelectron impact ionization. The main photochemical region of the ionosphere is dominated by two main layers: the so-called M2 at about 130 km formed by $O_2^+$ and $O^+$ above 250km (e.g. *Hanson et al. 1977; Benna et al. 2015*), and a second lower layer called M1 at about 115 km (e.g. *Peter, 2018*, naming convention after *Rishbeth and Mendillo (2004)*). A typical ionospheric profile for day and nightside is shown in *Figure 5*, together with several internal and external forcings such as solar radiation, meteors, electron and solar storm particle precipitation, solar wind, magnetic fields, gravity waves, dust storms, and atmospheric cycles.

The M2 peak density and altitude are known to be highly variable, depending on the solar flux, the solar zenith angle and the state of the underlying neutral atmosphere. A summary of the observed variability of the dayside ionosphere is given in *Withers (2009)*. The altitude of the M2 peak for a given solar zenith angle appears at approximately unity optical depth for EUV photons, which is approximately at a constant pressure level. The M2 altitude is therefore coupled to spatial and temporal variations in the underlying conditions in the lower neutral atmosphere (e.g. from planetary and tidal wave activity, *Bougher et al. (2017)*). Large amounts of Martian dust also affect the altitude of the M2 peak (*Wang and Nielsen 2003*), which distribution in years is quite irregular (e.g. *Montabone et al. (2015)*). Therefore, a **regular and frequent monitoring of the ionosphere of Mars is necessary to determine the global and localised effects of atmospheric dust on localized areas and for small time scales**. The fundamental formation mechanisms of the undisturbed dayside ionosphere of Mars are well understood. However, anomalous ionospheric shapes are regularly observed with electron density radio occultations profiles (*Withers et al. 2012*) **whose spatial/temporal extent, characteristics and origins are still under discussion.**

Regarding the lower secondary layer (M1), its altitude range is currently based only on the ionospheric electron density observations provided by the radio occultation technique (see *Figure 8B*). **No observations of the ion composition are provided on a regular basis for the whole ionospheric region below the M1 peak (<~100 km).** Therefore the origin of the M1 shape variability remains unclear. Single radio occultation observations of the M1 layer indicate that this layer responds to solar flares in the same way as the E region of the terrestrial ionosphere (*Mendillo et al. 2006*). However, **the effects of solar flares or solar energetic particles on this ionospheric region have never been investigated for short time scales**. The composition of the lower nightside ionosphere remains also unknown (*Girazian et al. 2017*). This also includes the details of the nitrogen cycle at Mars below ~120 km altitude (see e.g. discussion in *Lefevre and Krasnopolsky (2017)*). Moreover, the very few observations above surface regions with strong crustal magnetic fields are still inconclusive (*Andrews et al., 2015; Peter, 2018; Gupta and Upadhayaya, 2019*), although seem to have different composition and structure than in non-crustal field regions (e.g., *Withers et al., 2019*).

In 2005, *Pätzold et al. (2005)* discovered a local and sporadic third layer below the established two layered structure in the observations of the Mars Express MaRS radio occultation experiment. The excess electron density can be detached (Md, *Figure 6a*) or merged (Mm, *Figure 6b*) with the main ionospheric body. However, we do not know the ions that formed that layer because **no regular in-situ observations of the atmospheric and ionospheric composition have been conducted in the altitude region between 70 and 110 km.** Therefore the origin and composition of these **features remains unknown**. The observed excess electron density has been investigated by several modellers and attributed to the influx of meteoroids (e.g. *Molina-Cuberos et al.*

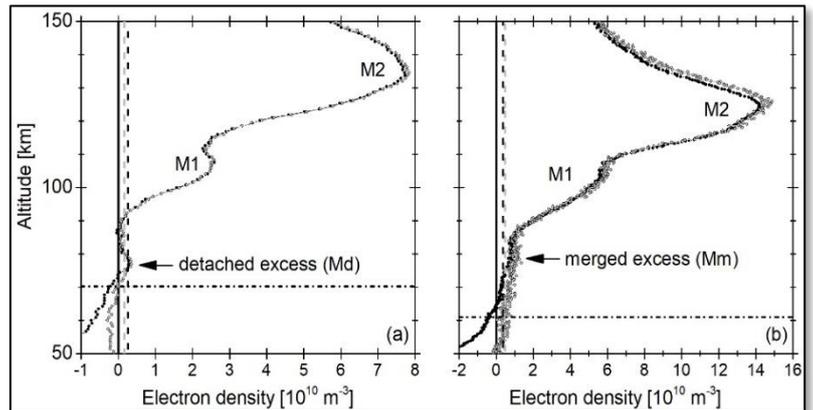

**Figure 6:** *MEX-MaRS dayside electron density. The black straight line is the zero line, the dashed black and gray lines indicate the associated noise levels, the black dash-dotted line is the lowest valid altitude of observation.. (a) Day of Year (DoY) 350 (2005), SZA = 74.07°. (b) DoY 337 (2013), SZA = 57.16°. Figure from Kerstin Peter, Universität zu Köln.*

*(2008), Whalley and Plane (2010))*. However, **due to a missing monitor for interplanetary dust particles at Mars, the meteoroid input flux for the models is poorly constrained**. The remote observation of meteoric $Mg^+$ by the MAVEN Imaging Ultraviolet Spectrograph (IUVS) on the planetary dayside indicated, however, that the permanently available layer of $Mg^+$ at ~75 km is too small to be the only source responsible for the identified excess electron densities below 110 km altitude with radio occultation. However, the remote MAVEN IUVS observations of $Mg^+$ are limited to above 75 km altitude on the planetary dayside (*Crismani et al. 2017*). The lack of a layer of neutral Mg below the identified layer of Mg+ (seen at Earth and predicted by most meteoric models for Mars) **challenges current models of the interaction between meteoroid material and the planetary atmosphere and ionosphere** (*Crismani et al. 2017; Plane et al. 2018*).

### 2.3.2. Does plasma reach the Martian surface?

Below the secondary ionospheric layer (M1 peak), it is believed that the ionosphere is still present **but no measurements of this region are available**. Some indirect observations, for example from the lack of reflected signal from the surface with radar soundings, indicate that low altitude ionization is present on the dayside, and also on the nightside when solar storms hit the planet. In those cases, low altitude ionospheric layers absorb the radar signals due to a high rate of neutral-electron collisions (e.g. *Němec et al., 2014; 2015; Sánchez-Cano et al., 2019*). In addition, the flux of galactic cosmic rays is being measured by the Mars Science

**Mars' plasma system**

Laboratory mission at the surface (e.g. *Guo et al., 2015*), and some works have theorized the effect of these galactic cosmic rays on the very low atmosphere creating an ionized layer at ~25-30 km altitude with a peak concentration of ~$10^9$ electrons per m$^{-3}$ depending on solar activity and aerosol formation close to the surface (e.g. *Whitten, et al., 1971*; *Haider et al., 2015; Grard, 1995; Cardnell et al., 2016,* see also *Figure 5*). This radiation, together with the solar UV photons also reach the surface of the planet, and are believed to ionise the neutral atmosphere and the aerosols closer to the ground forming positive ions, electrons and photoelectrons and **generate an electric field from the ground to the atmosphere** (*Grard, 1995; Cardnell et al., 2016*). Consequently, electric fields could be enhanced by the charged dust of the surface, especially at the dust seasons, having strong effects on the atmospheric conductivities, and therefore, on the ionosphere. **This is a totally unexplored region that requires a systematic exploration from the ground in conjunction with orbiter observations.**

### 2.3.3. Quantitatively, what is the role of lower atmospheric effects on the ionosphere?

The neutral atmosphere is responsive to topographic and temperature variations that occur diurnally, episodically, as well as seasonally (dust storms). The energy produced by such drivers produce gravity waves that propagate upward with altitude from the surface and are observed in neutral atmospheric observations. The ionosphere is generated from the neutral atmosphere, and plasma structure is, therefore, also reflective of this energy deposition. **The energy budget in the atmosphere of Mars remains unsolved** (*Matta et al., 2014, section 2.1.4*), and this investigation would be one of the **key pieces of this puzzle**.

The structure of the ionosphere of Mars is an excellent monitor for ambient dynamical processes. Upper atmospheric disturbances can produce structural variations in the upper atmosphere and lower atmospheric disturbances can propagate upward to reflect on plasma structure as well. The effects of gravity waves on the Martian ionosphere have been investigated to show non-negligible effects on atmospheric variability (*Yigit et al., 2015, England et al., 2019*). The ionosphere is closely coupled to the neutral atmosphere at altitudes where gravity wave perturbations are highly dynamic (*Mayyasi et al., 2019*).

Dust activity in the lower atmosphere results in significant oxygen depletions in the thermosphere.

Oxygen is the primary mediator of Mars's ionospheric photochemistry cycle, converting the primary ion $CO_2^+$ into the dominant ion $O_2^+$. When $O_2^+$ recombines with electrons, it dissociates providing energy for hot oxygen atoms to escape; this process has been the dominant source of escaping oxygen in recent times (*Lillis et al., 2017*). Dust storms are a special and characteristic form of dust activity at Mars, which are highly dynamic events that result in a strong upper atmosphere variability. A good example is the 2018 planet-encircling dust event (PEDE) that lasted a few terrestrial months, and whose effects on the upper atmosphere are still being analysed. Changes in circulation patterns and water propagation cycle at Mars due to dust storms are currently being investigated to determine how the lower atmosphere and upper atmosphere are linked. Dust storms can cause an upwelling of lower atmospheric species, such as water vapour, subsequently resulting in variations in the upper atmospheric composition (*Heavens et al., 2018*).

**The effects of lower and mid atmospheric variations on the upper atmosphere have yet to be quantified** due to the challenges of making in-situ lower and mid atmospheric measurements. Synoptic monitoring of lower atmosphere dust loading, middle atmospheric water abundance, and upper atmospheric hydrogen and oxygen response, as well as the temperature structure at all altitudes across multiple dust events, is required to **understand the processes (currently unknown) by which the lower atmosphere drives the upper atmosphere and escape.** Future missions should consider making **routine measurements of lower altitudes to close this essential gap in our knowledge** of the Martian atmosphere.

### 2.3.4. To what extent does the ionosphere permit and inhibit radio communication at the surface?

One of the consequences of having a thin atmosphere and being unprotected by an intrinsic global magnetic field is that the amount of particles (both from the solar wind and meteors) that precipitate into the Martian atmosphere is very large. These particles are known to produce ionization at low altitudes (below ~100 km) where the neutral atmosphere is denser and collisions are more common. Consequently, radio frequency absorption in the lower ionosphere is one of the most common phenomena that occur, which affects high-frequency (HF) operations and communications with and within the surface platforms. In contrast with Earth where HF malfunctions last of the order of few hours, at Mars these issues typically last on the order of several days (and even weeks). These phenomena make future human exploration challenging. For example, meteor and cometary dust showers are a well-known source of ionospheric absorption at low altitudes (*Molina-Cuberos et al., 2003; Gurnett et al., 2015; Crismani et al., 2017*). However, the most challenging phenomena, in terms of scientific exploration and instrument operations, are Space Weather events. SEPs are the most intense source (both in length and in reaction time) of ionization at low altitudes. It has been long known that SEPs are able to produce large malfunctions in HF operations, such as total radar blackouts (e.g. *Espley et al., 2007*). However, the type of particles and the mechanisms behind those blackouts were unknown. Recently, the two radars that are currently working in Mars' orbit and sounding the ionosphere, surface and subsurface of the planet suffered a complete radio blackout during a large SEP event in September 2017, i.e. MARSIS (Mars Advanced Radar for Subsurface and Ionosphere Sounding) on board Mars Express and the Shallow Radar (SHARAD) onboard Mars Reconnaissance Orbiter (MRO). *Sánchez-Cano et al. (2019)* in line with *Ulusen et al. (2012)* analyses found that high-energy electrons accelerated by the solar wind created a dense and global layer of ions and electrons at ~90 km around the whole planet. This layer attenuated radar signals continuously for 10 days, preventing the radars from receiving any signal from the planetary surface. The main properties of the low ionosphere was estimated using a combination of data analysis from the MAVEN, Mars Express and MRO orbiters together with numerical simulations of the ionospheric response. This is only an indirect low-limit estimation of the low ionosphere properties because the **low ionosphere (in the mesosphere region) has never been explored**.

Understanding the Martian response to Space Weather is essential in order to assess how the plasma environment reacts and dissipates energy from the solar storms. This includes understanding how common these absorption layers are, the nature of their vertical structure (and if they reach the surface of the planet under certain conditions), their local time variation, and their lifetimes. Moreover, **understanding how low atmospheric layers affect the communications will help us to improve technology, as well as mitigate the risk for human and robotic exploration missions.**

In addition, the ionosphere has strong effects on radio propagation due to electromagnetic dispersion within the ionospheric plasma. This is a well-known problem for the MARSIS and SHARAD radars that sound the surface of Mars (e.g., Sanchez-Cano et al., 2015a), but also, for potential orbital network of communications and navigational satellites at Mars (Mendillo et al., 2004). **A**

**good understanding of the ionospheric-induced scintillations and group delay effects is certainly a capability needed for human exploration of the red planet, because they have the potential of affecting the fundamental goal of a GNSS-type system at Mars.**

### 2.3.5. What role do winds play on wave propagation?

The dynamics of the thermosphere are dominated by atmospheric wave activity at both global (tides) (*Liu et al., 2017, England et al., 2016*) and small scales (gravity waves) (*Yiğit et al., 2015*). These waves impact the dynamics, energetics (temperature structure) and even composition of this region, all of which have subsequent influences on atmospheric escape. The character of these waves appears to change as they move from the well-mixed atmosphere below 100 km to the diffusion-dominated region in the thermosphere. However, despite many missions sampling the thermosphere in situ (e.g. Mars Express, ExoMars, MAVEN), **much is unknown**, as for example the altitude of this transition, how it occurs, or what the true impact of these waves are.

**The nature and impact of these waves is not understood because we do not have a coherent picture of the winds in the Martian thermosphere.** The limited set of direct measurements of the winds from MAVEN-NGIMS (*Mahaffy et al., 2014*) from ~140 to 240 km orbit-to-orbit changes of 100-200 m/s, which are as large as the mean winds themselves. These observed variations cannot be explained by current atmospheric models. The role that atmospheric waves play in producing such variations remains unknown and requires systematic measurements of these winds simultaneous with density structures, rather than the short, isolated campaigns.

### 2.3.6. What are the roles of small scale ionospheric irregularities and electrodynamics in the Martian ionosphere?

Thanks to the well-equipped plasma package on the MAVEN mission, we have recently discovered the existence of small-scale ionospheric irregularities in the Martian ionosphere (*Fowler et al., 2017a*), which are assumed to be stationary. These irregularities are characterized by quasi sinusoidal variations in the magnetic field strength at length scales of 5-20 km perpendicular to the local magnetic field, and accompanied by large variations in the ionospheric electron density. These irregularities are observed primarily in the Martian dynamo region of the ionosphere (~130–170 km altitude) at specific local times, solar zenith angles and planetary latitudes and longitudes, during conditions when ions are unmagnetized due to frequent collisions with the neutral atmosphere, but electrons remain magnetized (*Fowler et al., 2019*). Such irregularities have been studied extensively at Earth since the 1930s (*Fejer and Kelley, 1980; Kelley and McClure, 1981; Ossakow, 1981*) and are generated there primarily by the gradient drift and two stream instabilities at the magnetic equator where the magnetic field is horizontal. The study of ionospheric irregularities at Earth has provided a wealth of information related to the coupling between the thermosphere, ionosphere and terrestrial dipole magnetic field, including the local and global current systems that arise from these couplings. Evidence for strong ion-neutral coupling has also been recently demonstrated using simultaneous observations from the MAVEN mission (*Mayyasi et al., 2019*), and shows that the neutral atmosphere is a significant driver of this plasma structure as can be seen in e.g. *Figure 7*, where the same variability observed in the argon profiles is clearly seen in the main ions and electron profiles.

Contrary to Earth, **the study of Martian ionospheric irregularities is still in its infancy** because spacecraft at Mars prior to MAVEN were in orbits that did not sample the dynamo region of the ionosphere, or did not possess instrumentation capable of observing such irregularities. As a result, there is still much that is unknown, including which ionospheric instabilities are responsible for their generation. Plasma instrumentation carried by MAVEN is unable to resolve the density irregularities due to relatively long measurement integration times and cadences, and it is not known whether the density and magnetic field variations occur in or out of phase of each other (or perhaps neither). Ionospheric density measurements that are able to resolve these 5-20 km length scales, and whether they are stationary or not, would enable a quantitative characterization of the density variations and their relation to the magnetic field variations. Electric field fluctuations are also associated with terrestrial ionospheric irregularities and can be used to characterize the dominant wave numbers of the observed irregularities (e.g. *Fejer and Kelley, 1980*). Such measurements are limited in the Martian ionosphere because MAVEN's Langmuir Probe and Waves (LPW) instrument provided one dimensional electric field wave power spectra throughout 2015 only (*Fowler et al., 2017b*). Acquiring even one dimensional time series electric field data during such irregularity events at Mars would greatly aid in conclusively identifying which instabilities are responsible for the generation of the Martian ionospheric irregularities. In the terrestrial ionosphere, currents driven by strong ion-neutral coupling (while electrons remain magnetized) can be important drivers of these irregularities (e.g. *Oppenheim, 1997*). The Suprathermal and Thermal Ion Composition (STATIC) instrument on MAVEN is capable of measuring ion winds under specific ionospheric conditions, but the caveats and limitations of these measurements mean that typically only the cross track ion wind velocity is measurable during irregularity events. Uncertainties on these measurements can be somewhat large, around 100 m/s, which at times can be almost as large as the background cross track wind velocity. Three dimensional ion or neutral wind measurements, at cadences of 5-10 s and uncertainties <50 m/s, would greatly aid in determining the role that ion-neutral winds play in the formation of these irregularities at Mars.

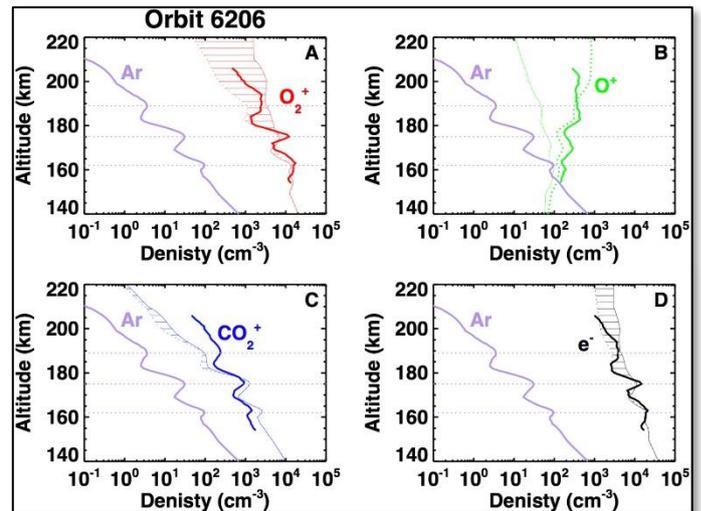

**Figure 7:** *MAVEN plasma density profiles from orbit 6206. Simulations with photochemistry are shown as thin dotted profiles, and simulations with added transport are shown as thin solid profiles. This figure highlights the large role that the neutral atmosphere has a driver for small plasma structure. Figure from Mayyasi et al., (2019).*

**A whole host of comparative aeronomy questions also remain unanswered.** Examples include understanding how the different magnetic environments at Earth and Mars influence the formation of ionospheric irregularities. While Earth's ionosphere is dominated by the dipole magnetic field, Mars' magnetic environment is highly variable in both time and space due to the crustal magnetic fields that rotate with the planet (*Figure 2*), and the nature of the induced magnetosphere (*Figure 1*) that is highly responsive to changes in the upstream solar wind. The formation of ionospheric irregularities at Earth show strong seasonal dependencies (e.g. *Arras et al., 2008*). The precession of MAVEN's orbit means that currently, the MAVEN dataset does not provide enough coverage in time, planetary longitude, latitude, solar zenith angle or local time, to conclusively determine if seasonal variations exist or not. **A dataset spanning several Martian years and simultaneous multi-point observations is required to conclusively determine if seasonal dependencies exist there.**

### 2.3.7. How do low atmospheric cycles affect the upper atmosphere and escape?

Several lower atmosphere mechanisms are known to have an influence on the upper atmosphere, e.g. gravity waves, crustal magnetic fields, etc. Evidence suggests that lower and upper atmospheres of Mars are more closely connected than previously realized, affecting both hydrogen and oxygen escape. This seems partially caused by the absence of stratosphere at Mars. Exospheric hydrogen density (and associated Hydrogen escape) was observed to be strongly responsive to season (*Clarke et al., 2014; Chaffin et al., 2014; Bhattacharyya et al., 2015*), with the highest escape rates in southern summer. Middle atmospheric water abundance, which responds strongly to dust events (*Fedorova et al., 2018; Vandaele et al., 2019*), is also correlated with maxima in Hydrogen loss to space (*Heavens et al., 2018*).

Moreover, low-mid atmospheric cycles, such as the water and $CO_2$ cycles, have been recently discovered to have a very notable influence on the upper atmosphere (*Sánchez-Cano et al., 2018a*). The ionospheric total electron content acts as a perfect tracer for the thermosphere, which itself is affected by low-mid atmosphere variations. An example is the $CO_2$ cycle that results in the mass of the atmosphere varying by up to 30% every Martian year due to the polar caps' sublimation. The routine ionospheric observations from Mars Express appear to be an excellent indicator of the dynamic of this coupling, which is especially notable at northern spring as corroborated by observations from the SPICAM instrument onboard Mars Express and the REMS instrument onboard the Mars Science Laboratory, and modelling (*Sánchez-Cano et al., 2018a*).

However, **all these connections need to be understood, especially when other major internal drivers such as global dust storms significantly modify all these forcing every Martian year**. Simultaneous atmospheric observations (density, temperature, dust opacity, etc.) at different altitudes and on the same location are needed in order to understand the chemistry and physics of the lifting mechanisms and couplings and between different atmospheric layers, and their effect on seasonal atmospheric escape.

## 2.4. SCIENCE QUESTION 4

### Why should we have a permanent in-situ Space Weather monitor at Mars?

Space weather real-time forecast at Mars is currently very challenging because among other factors, it needs a continuous solar wind monitoring platform to provide timely and accurate Space Weather information. **This is only possible if sufficient observation data are continuously available.** At Earth, we have several spacecraft that for a few decades have been monitoring the Sun's activity and the solar wind. In fact, the most possible accurate measurements of the upstream solar wind at Mars occur when Mars and Earth are in apparent opposition or perfectly aligned in the Parker spiral (once every ~two years) because plasma missions such as Mars Express or MAVEN do not continually sample the solar wind. The Mars Upper Atmosphere Network (MUAN) community (*Opgenoorth et al., 2010*) has been leading coordinated efforts to have several Mars Express campaigns (with as many plasma instruments operating as possible) when both planets were aligned along the Parker spiral to better understand any Martian plasma variability due to external conditions (*Opgenoorth et al., 2013*).

However, the main problem arises when both planets are not close to each other, which happens for about a (terrestrial) year and half. In those situations, Mars does not have a permanent in-situ solar monitor and the analysis of several Space Weather effects on the Martian environment can be extremely difficult as they depend on solar wind observations taken in the best of the cases few hours before when the spacecraft was in the solar wind. The arrival of MAVEN in 2014 has improved our capability to monitor solar activity, in part due to its comprehensive aeronomy instrumentation suite. However, MAVEN still does not sample the solar wind 100% of time, meaning that assumptions and proxies must be used during time periods where solar wind observations are not present. MAVEN is providing additional contextual information of the near-Mars Space Weather disturbances, including their solar and heliospheric sources (*Lee et al., 2017*). Since 2014, there have been several coordinated efforts between Mars Express and MAVEN teams to have solar wind observations from one spacecraft while the other one takes upper atmosphere observations. However, the orbital period of MAVEN changed in 2018 after an aerobraking campaign, having the orbit's apoapsis reduced. As a consequence, MAVEN is now taking less in-situ solar wind data than before.

As largely discussed in this White Paper, continuous in-situ solar wind and Space Weather observations are extremely important for most of the science questions that still remain unknown at Mars. **A continuous in-situ solar wind monitor at Mars, together with atmospheric simultaneous observations is a first need** in order to fully understand the 3D dynamics of the plasma system, as well as for having an efficient and continuous thermosphere – ionosphere – magnetosphere –solar wind monitoring service which is absolutely needed in the eve of the Martian human exploration. This is perhaps more important at Mars than at Earth from the purely science point of view because Mars does not have a global intrinsic magnetic field that partly shields the planet like in the Earth's case. Therefore, Space Weather activity has a more dominant role in most of the Martian upper atmospheric processes that we have discussed, as well as on the amount of radiation that reach the surface of the red planet (e.g. *Guo et al., 2015*).

**The ideal situation for the next generations would be to have continuous Space Weather monitors at different Solar System positions**, in order to have efficient forecasting tools at different planetary environments, as well as to better understand the evolution of the Space Weather events. Moreover, we emphasize the importance of a Space Weather monitoring package, including a magnetometer, to be embarked in all planetary and astronomical missions as a basic payload requirement as discussed in *Witasse et al. (2017)*, as well as have the plasma instruments in continuous operation during solar superior conjunctions, even if only at a very low data rate, or continue to acquire data for later download.

## 3. Mission Concepts

In this section, we develop complementary concept ideas for the next generation of Mars' exploration based on coordinated multi-point science from a constellation of orbiting and ground-based platforms which focus on understanding and solve the current science gaps. The proposed missions could fit into an M-class mission. With the used of these type of mission concepts, we will be able to answer the science questions discussed in this White Paper, and get a global understanding of the 3D structure of the Martian plasma system, atmospheric coupling (from the surface to space), and solar wind driven ionosphere dynamics. Coordinated multi-point observations have the scientific potential to track these dynamics, and they constitute the next generation of Mars' exploration. *Table 2* summaries the type(s) of mission(s) that would be ideal to address the science questions described above.

### 3.1 Multi-satellite approach

The four main science objectives described in this White Paper can be addressed with a multi-satellite approach. There are different orbital configurations that can be considered with similar benefits as are discussed in the following. In all the configurations, a spacecraft that continuously samples the undisturbed solar wind at Mars is crucial, while the other(s) takes observations within the Martian system.

#### 3.1.1. A mothership with a network of small satellites

The most ideal scenario to address the four main science goals *(see Tables 1 and 2)* is to have a mothership on a slightly elliptic orbit near Mars dedicated to take measurements of the Martian ionosphere and upper atmosphere, while a network of small satellites (or even nano-satellites) are dedicated to different tasks, such as the monitoring of the solar wind, and characterization of the induced magnetosphere and lower atmosphere. **Ideally, 4-satellites measurements are the only way to unambiguously disentangle spatial and temporal variations** and compute currents, plasma wave, boundary crossings, and velocities providing that the spacecraft are close enough with respect to the plasma microscopic scales like inertial lengths and gyroradii. This concept idea has been already proposed to both ESA and NASA space agencies with some slightly differences in the configuration by *Leblanc et al. (2018)* and *Lillis et al., (2019)*, respectively, as it offers the most complete exploration of the whole Martian plasma system.

The mothership should be a traditional large (> 1000 kg) spacecraft well-equipped with atmospheric and plasma instrumentation capable of measuring many upper and lower neutral atmospheric variables (e.g. winds, pressure, temperature, aerosols, $H_2O$, etc.) precessing in local time. The mothership should have an elliptical orbit with periapsis at ~150 km and apoapsis at 5000 – 7000 km to accommodate multi-point plasma measurements.

*Table 2: Techniques and payload to address the Science Objectives*

| Science Objectives | Mission-type concept | Fundamental payload | Important payload |
|---|---|---|---|
| **SCIENCE QUESTION 1** How does solar wind driving impact on magnetospheric and ionospheric dynamics? **SCIENCE QUESTION 2** What is the structure and nature of the tail of Mars' magnetosphere at all scales? | a) Constellation of several nanosatellites and a mother spacecraft b) Constellation of two orbiters: one spacecraft placed on the upstream solar wind and the second spacecraft have a much longer orbital period to allow transit the further tail c) Use Phobos and Deimos as travel platforms | - Magnetometer<br><br>- Ion mass spectrometer (able to resolve at least $H^+$, $He^+$, $O^+$, $O_2^+$, $CO^+$)*<br><br>- Electron spectrometer*<br><br>- Langmuir Probe*<br><br>- Energetic particle detector (electron and protons)<br><br>- EUV monitor in all wavelengths* | - Ionospheric radar (topside and bottomside) - Neutral mass spectrometer* - Energetic Neutral Analyser* - Radiation monitor - Neutron monitor - Electric field - Wind interferometer - Radio-occultation with Earth and between satellites* - VHF TEC instrument - IR and UV spectroscopy* - LIDAR |
| **SCIENCE QUESTION 3** How does the lower atmosphere couple to the upper atmosphere? | a) Constellation of several nanosatellites and a mother spacecraft b) Dual radio-occultations between two orbiters (related to *a)*) c) Ionospheric sounding from above and below d) Remote sensing atmospheric instrumentation in orbit | | |
| **SCIENCE QUESTION 4** Why should we have a permanent in-situ Space Weather monitor at Mars? | a) Constellation of several nanosatellites and a mother spacecraft: An orbiter placed always on the upstream solar wind b) Use Phobos and Deimos as travel platforms | | *only for orbiters* |

The other small spacecraft should be devoted to different tasks, such as for example:

- An orbiter dedicated to the monitoring of the solar wind to be placed in a large circular orbit in the upstream solar wind. It could be also at areostationary orbit (>10,000 km altitude).
- Two (or more) polar-orbiters spaced in local time and monitoring a subset of lower atmosphere variables (e.g. temperature, aerosols), some at lower fidelity (e.g. $H_2O$).
- Two identical spinning orbiters dedicated to the characterization of the induced magnetosphere and the far tail and for electric field measurements.
- Two orbiters on an areostationary orbit, spaced equally in longitude, enabling complete diurnal and geographical coverage up to ~70° north and south latitudes and views of the hydrogen and oxygen exospheres.

All spacecraft except the mothership are expected to be small (< 100 kg) satellites, capable of direct communication with Earth but primarily using the mothership to relay their data back. This type of mission could be done with international collaboration, where one country has the major role controlling of the mission and build the mothership and different countries/agencies built the other small spacecraft.

An important aspect of the Martian plasma system that can be systematically explored with a mission of this type is the bottomside structure of the ionosphere, as well as the coupling with the lower atmosphere via **dual radio-occultations between all the spacecraft network** (e.g. *Ao et al., 2015*). The dual radio-occultation technique provides a measure of the electron density along the line of sight between both spacecraft (*Figure 8B*). In-situ dual radio-occultations provides a much better coverage of the planet with respect to local time as compared to the typical occultation using Earth as receptor because currently, only solar zenith angles larger than 45º can be sampled due to geometric limitations between both planets. Also, it will reduce the error of the retrievals as the signals do not need to cross the space and Earth's ionosphere. As an additional advantage, there is no need for a proper instrument as the communication system between the different spacecraft can be used to perform the radio-occultations.

### 3.1.2. Twin orbiter constellation

Another feasible scenario to accomplish most of the science goals is to have a twin orbiter constellation precessing in local time in where one spacecraft has a near circular orbit (or low elliptical orbit) with apoapsis at 5000-7000 km (outside the Martian bow shock) to be able to monitor the solar wind, and the other one has an elliptical orbit with long period to be able to transit the far Martian tail. The period of both orbits should account for the largest possible amount of time of one of the twin spacecraft being on the upstream solar wind. The physical characteristics and instrumentation of both twin spacecraft should be similar to the mothership described in *Section 3.1.1*, and also able to perform dual radio-occultations when location-wise possible. Both spacecraft could have either polar or equatorial orbits. For the one with shortest period, it would be recommended to have a polar-orbiter precessing in time. For the orbit with longest period, both types of orbits are adequate giving precious information on the 3D structure of the nightside magnetosphere and tail. However, an equatorial orbit would be perhaps more adequate to study the different structures of the tail, including the width of the plasma sheet and magnetosheath, and to calculate the total amount of ion outflow and currents through the tail as would provide the whole horizontal structure of the tail in every orbit transit.

### 3.1.3. Phobos and Deimos as travel platforms

The near equatorial and circular orbits of the Martian moons Phobos and Deimos offer also great opportunities **for long-life and low maintenance stations**, as recently assessed by *Sefton-Nash et al. (2018)*. Moon stations could be used for different purposes such as for meteorological studies, or data relays, but also, for plasma physics. The two moons, Phobos and Deimos, orbit Mars at ~3 and ~7 Mars radii respectively, and cross the whole horizontal structure of the tail (including bow shock boundary) several times per day, as well as transit the solar wind in each orbit. Both moons offer large possibilities for science, such as evolution of the solar wind between 7 and 3 Mars radii and the bow shock, or one moon being a solar wind monitor and the other one been sampling the tail. However, specifically the orbit of Deimos would enable studies at larger distances from Mars.

### 3.1.4. Payload to consider

The payload that each spacecraft should carry will depend on the different science scenarios considered before. For example, in the case of the twin spacecraft, the same instrumentation should be considered for both of them. However, in the case of the mothership and small satellite network scenario, the distribution of the payload would depend on the objectives of each satellite. *Table 2* gives also an overview of the basic payload that should be considered, despite the format of the mission, including the most fundamental instruments that should be always included and keep in operation, such as a magnetometer, ion and electron electrostatic analyzers, a Langmuir probe for ionospheric densities and temperatures, an energetic particle detector and a solar EUV monitor. The rest of the instruments considered as "important" are also essential to address the Science Objectives, although in less extent that the fundamental payload.

## 3.2. Ground-based network approach

In order to determine the vertical ion and electron distribution of the bottomside Martian ionosphere (from the surface to the main ionospheric peak), ground-based ionospheric measurement techniques are also feasible for Mars. The extremely low conductivity of the arid Martian surface is indeed favourable for such systems as simple, lightweight antennas can be deployed directly onto the surface without negatively affecting their performance. Ground-based measurements of the ionosphere have been proposed in the past (*Berthelier et al., 2003*) but no such system has yet flown to Mars. Conceptually, there are two simple ways in which we could retrieve more information of the low ionosphere of Mars: a) Systematic use of radio-occultation between different spacecraft and/or ground detectors (e.g. *Ao et al., 2015*). b) Having a network of digisondes on the surface of Mars able to systematically sample the bottomside ionosphere.

Focusing on the ground-based network, two measurement techniques can be considered. The first one is a relative ionospheric opacity meter (Riometer), which operates via the passive measurement of ionospheric attenuation of cosmic radio sources at 0.1 to 35 MHz frequencies. While this technique does not give information about the vertical structure of the ionosphere, it provides a useful counterpart to orbital measurements, by accurately constraining the diurnal variation of the ionosphere at a fixed location on the surface. The second one is a more complex active ionospheric radar experiment (Ionosonde), comparable to the MARSIS instrument (*Figure 8A*) onboard Mars Express. Such an instrument operates by transmitting short radio pulses at a range of frequencies, which reflect from the ionosphere at different altitudes, and measuring the delay time before they are again received back on the same antenna. In this way, a full profile of the plasma density variation with altitude is obtained for the bottomside ionosphere (see *Figure 8C*). Indeed, both systems can be based around a single dipole antenna and shared electronics. In order to achieve sufficient performance at frequencies at and below ~1 MHz, at least a dipole antenna of length >10 m is required. The low transmitted powers required, chemically inert environment and low pressure exerted by even "strong" winds on Mars allow for a very lightweight antenna design. Deployment of a large antenna on the surface from a stationary platform requires further study. A range of technical solutions can be conceived; 'dragging' an antenna onto the surface using an accompanying rover, spring-loaded deployment or pyrotechnic deployment using small rockets, or perhaps even using an inflatable antenna structure. For the case of a dipole antenna, while the ideal situation is deployment in a perfectly straight line, the performance is highly tolerant of even large departures from this. Likewise, most of the Martian surface is suitable for such instrumentation. The extremely low conductivity of the surface and sub-surface at the relevant frequencies is favourable for such a system.

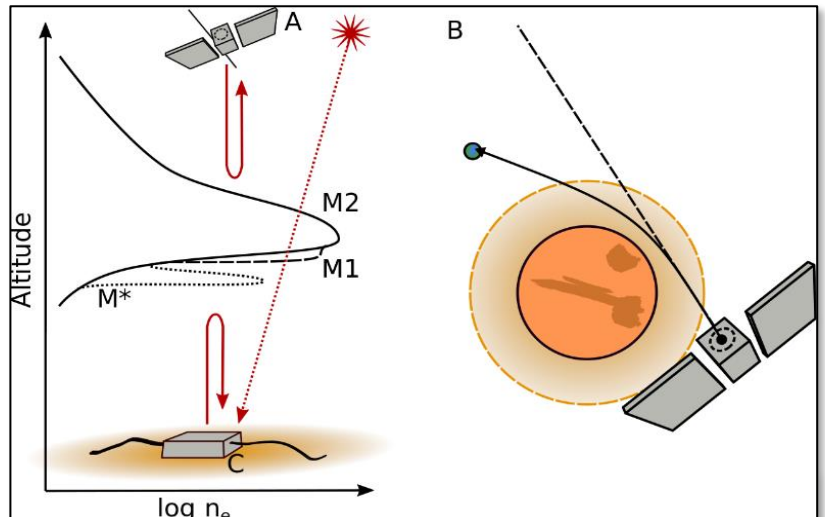

**Figure 8:** *Left: Mars' electron density profile, including the M2 and M1 layers, and a transient lower layer M\*, and schematics of the detection of this profile using orbital sounding (A), surface sounding (C) and (right) radio occultation (B). Figure from David Andrews (IRF Uppsala).*

In addition, a stationary surface science platform could well also make measurements of the local magnetic field variations associated with ionospheric currents (e.g. *Lillis et al., 2019*). In fact, if **every rover or surface platform that it is sent to Mars in the near future provides a magnetometer like Insight, our knowledge of the surface-magnetosphere coupling via ionospheric currents would be further advanced**.

## 4. Conclusions

The future of the Martian science and exploration requires **coordinated multi-point plasma measurements with high temporal resolution** to be able to untangle the whole Martian dynamic system, from its surface until space. This is extremely important for a good comprehension of the Martian system as a whole, but also to understand the real variability of unmagnetized bodies. We have now **a unique opportunity at Mars** to perform comparative planetology science (and extrapolate knowledge to other bodies and solar systems), as Mars is the only body beyond Earth where this type of exploration can be currently done.

We have identified **four main science questions** that are currently unanswered at Mars (see *Table 1*), which are related to dynamic process at the dayside and nightside magnetosphere and ionosphere, as well as coupling with the lower atmosphere and surface. In particular, there are still **two important observational gaps** in the Martian system that no mission has been able to fully explore: the 3D structure of the full Martian tail and its dynamics, and the lower Martian ionosphere from the surface until ~80 km which need to be solved. To resolve all these science questions, **there is also a clear need for an efficient solar wind monitor at Mars**.

Finally, two mission concepts are also discussed based on coordinated multi-point science from a constellation of orbiting and ground-based platforms, which focus on understanding and solving the current science gaps.

# Science Team

1. **Beatriz Sánchez-Cano**
   *University of Leicester, United Kingdom*

2. **Mark Lester**
   *University of Leicester, United Kingdom*

3. **David J. Andrews**
   *Swedish Institute of Space Physics Uppsala, Sweden*

4. **Hermann Opgenoorth**
   *Umeå University, Sweden*

5. **Robert Lillis**
   *University of California Berkeley, United States of America*

6. **François Leblanc**
   *Laboratoire atmosphères, milieux, observations spatiales / Centre national de la recherche scientifique, Sorbonne Université, France*

7. **Christopher M. Fowler**
   *University of California Berkeley, United States of America*

8. **Xiaohua Fang**
   *University of Colorado Boulder, United States of America*

9. **Oleg Vaisberg**
   *Space Research Institute of Russian academy of Sciences, Russia*

10. **Majd Mayyasi**
    *Boston University, United States of America*

11. **Mika Holmberg**
    *Institut de Recherche en Astrophysique et Planétologie, France*

12. **Jingnan Guo**
    *University of Science and Technology of China, China Christian-Albrechts-Universitycity, Germany*

13. **Maria Hamrin**
    *Umeå University, Sweden*

14. **Christian Mazelle**
    *Institut de Recherche en Astrophysique et Planétologie, France*

15. **Kerstin Peter**
    *Rheinisches Institut für Umweltforschung an der Universität zu Köln, Germany*

16. **Martin Pätzold**
    *Rheinisches Institut für Umweltforschung an der Universität zu Köln, Germany*

17. **Katerina Stergiopoulou**
    *Swedish Institute of Space Physics Uppsala, Sweden*

18. **Charlotte Goetz**
    *Technische Universität Braunschweig, Germany*

19. **Vladimir Nikolaevich Ermakov**
    *Space Research Institute of Russian academy of Sciences, Russia*

20. **Sergei Shuvalov**
    *Space Research Institute of Russian academy of Sciences, Russia*

21. **James Wild**
    *Lancaster University, United Kingdom*

22. **Pierre-Louis Blelly**
    *Institut de Recherche en Astrophysique et Planétologie, France*

23. **Michael Mendillo**
    *Boston University, United States of America*

24. **Cesar Bertucci**
    *Instituto de Astronomía y Física del Espacio, Argentina*

25. **Marco Cartacci**
    *Istituto Nazionale di Astrofisica, Italy*

26. **Roberto Orosei**
    *Istituto Nazionale di Astrofisica, Italy*

27. **Feng Chu**
    *The University of Iowa, United States of America*

28. **Andrew J. Kopf**
    *The University of Iowa, United States of America*

29. **Zachary R. Girazian**
    *The University of Iowa, United States of America*

30. **Michael T. Roman**
    *University of Leicester, United Kingdom*